\documentclass[12pt]{article}
\usepackage{graphicx}

\def \bea{\begin{eqnarray}}
\def \beq{\begin{equation}}

\def \eea{\end{eqnarray}}
\def \eeq{\end{equation}}

\begin{document}
\rightline{EFI 09-36}
\rightline{arXiv:0912.1053}
\rightline{December 2009}
\bigskip
\centerline{\bf THE MYSTERY OF PARITY\footnote{Chapter for Festschrift in
honor of Tom Erber's 80th birthday}}
\bigskip

\centerline{Jonathan L. Rosner\footnote{rosner@hep.uchicago.edu}}
\centerline{\it Enrico Fermi Institute and Department of Physics}
\centerline{\it University of Chicago, 5640 S. Ellis Avenue, Chicago, IL 60637}

\begin{quote}

And should I not take pity on Nineveh, that great city, with more than a
hundred and twenty thousand inhabitants who do not know their right
hand from their left, and many beasts besides?\footnote{Jonah 4:11}

\end{quote}

\section{Introduction}

Our world does not exhibit left-right symmetry at the level of familiar
objects:  biological \cite{Pasteur} and polar \cite{Erber} molecules, organic
chemicals \cite{Hoffmann}, human anatomy, and much more.  However, before 1956
it was widely \cite{Pais} (though not universally \cite{Farmelo}) assumed that
the fundamental laws of physics exhibited that symmetry ({\it parity
invariance}).  When it was called into question for the weak interactions
\cite{LeeYang}, experiments \cite{Wu,GLW,FT} quickly showed that in fact the
weak interactions had a definite handedness, involving left-handed particles
and right-handed antiparticles.

Could parity violation at the microscopic level be responsible for what we see
in the macroscopic world?  Despite calculations claiming this to be so
(see, e.g., \cite{Kondepudi}), Tom Erber has pointed out that this asymmetry
need not stem from the microscopic level, but can arise spontaneously in very
simple systems.  For $N$ point charges arranged on the surface of a unit
sphere, the lowest-energy state is mirror-symmetric for $2 \le N \le 10$, but
asymmetric for $N=11$ \cite{Erber,Coffey} and specific higher values of $N$.
In like manner (see also \cite{Gleiser}), although the Coriolis force tends to
send water down a drain counterclockwise in the Northern Hemisphere and
clockwise in the Southern, initial conditions play a far more important role.

In this article we shall be concerned with {\it microscopic} parity invariance
and a mystery which it presents.  This consists of a marriage of internal
and space-time symmetries, forbidden when the space-time symmetry
consists of the whole Poincar\'e group \cite{Coleman}
but permitted in this case because of the discrete nature of the parity
transformation.  We will argue that this marriage could point to regularities
underlying the nature of quarks and leptons, and to extensions of particle
interactions beyond those known today.

In Section II we briefly review the observed pattern of quarks and leptons,
noting the great difference between the masses of the light neutrinos and
the remaining fermions.  In Section III we express this difference in
group-theoretic terms, relying on an oft-employed five-dimensional geometric
construction based on the group SO(10).  The role of parity reversal in this
language is extremely simple, consisting of reflection of one of the five
coordinates.  Possible consequences of this observation are given in Section
IV, while Section V concludes.

\section{Quark and lepton patterns}

\begin{figure}[h]
\includegraphics[width=6in]{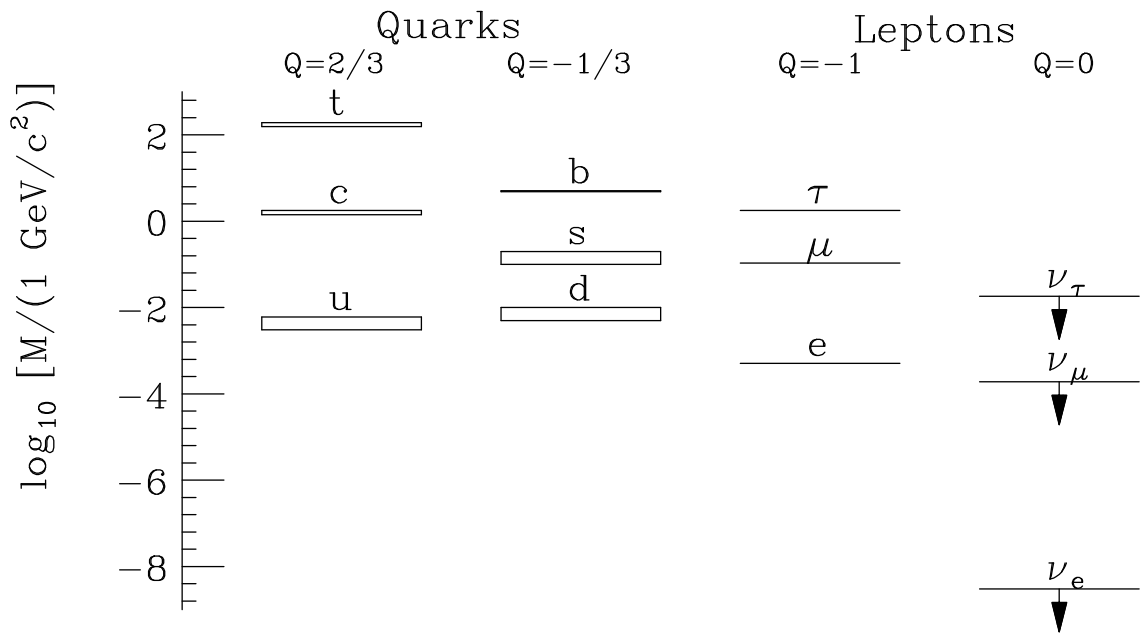}
\caption{Masses of observed quarks and leptons on a logarithmic scale.  The
upper limits on neutrino masses are based on direct searches \cite{PDG08};
see text for much more stringent limits.
\label{fig:qls}}
\end{figure}

The observed quarks and leptons fall into three families.  One distinguishes
left-handed from right-handed states.  Each left-handed family consists
of a quark electroweak doublet [transforming as a triplet of color SU(3)], a
lepton doublet [transforming as a singlet of color SU(3)], and the
corresponding antiparticles which are all electroweak singlets.  In each
right-handed family the roles of the particles and antiparticles are reversed.
For Dirac particles (the quarks and charged leptons) the left-handed and
right-handed states and the corresponding antiparticles are combined into
one four-component object with a specific ``Dirac'' mass.  The possibility
that a neutrino can be its own antiparticle allows for left-handed neutrinos
and right-handed antineutrinos (the ``active'' participants in weak
interactions) to have one ``Majorana'' mass while the ``sterile'' left-handed
antineutrinos and right-handed neutrinos have another.  In
Fig.\ \ref{fig:qls} we show present information on quark and lepton masses,
quoting Dirac masses for the charged fermions and direct upper limits on
masses of ``active'' neutrinos which may or may not be of Majorana type.

Neutrinos are known to mix with one another, so that the states of definite
mass (denoted $\nu_1$, $\nu_2$, and $\nu_3$) are linear combinations of
the ``flavor'' eigenstates $\nu_e$, $\nu_\mu$, and $\nu_\tau$.  Neutrino
oscillation experiments find $\Delta m_{21}^2 \equiv m_2^2 - m_1^2 = (7.59
^{+0.19}_{-0.21})\times 10^{-5}$ eV$^2$, $\Delta m_{32}^2 \equiv m_3^2 - m_2^2
= (2.43 \pm 0.13) \times 10^{-3}$ eV$^2$ \cite{PDG08}.  If $m^2_1 \ll
m^2_{(2,3)}$, then $m_1 \ll m_2
\simeq 9$ meV, $m_3 \simeq 50$ meV.  However, all the neutrino masses could
be larger and quasi-degenerate.  In any case a cosmological bound implies that
the sum of the (active) neutrino masses must not exceed 0.28 eV
\cite{Bernardis}, far below the direct limits depicted in Fig.\ \ref{fig:qls}.

Fig.\ \ref{fig:qls} represents one of the great puzzles of today's particle
physics.  Do the masses of the quarks and leptons (and their mixing under the
weak interactions) represent some deep underlying structure (as in the
Periodic Table of the Elements), or the solution of some anarchic dynamics
(as in the Titius-Bode law describing planetary orbits)?  For the present we
bypass this question and discuss the structure of a single family, which we
shall denote
\beq \label{eqn:fam}
F = \left( \begin{array}{c} u \\ d \\ \nu_e \\ e^- \end{array} \right)~.
\eeq

\section{Geometry of grand unified groups}

The strong interactions are described by an SU(3)$_C$ (C for color) Yang-Mills
gauge theory, while the gauge symmetry of the electroweak interactions is
SU(2)$_L \otimes$ U(1)$_{Y_W}$, where the subscript $L$ indicates that the
interaction applies to left-handed fermions (and right-handed antifermions),
while the subscript $Y_W$ denotes weak hypercharge.  Georgi and Glashow
\cite{GG} found an ingenious way to unify SU(3)$\otimes$SU(2)$\otimes$U(1) into
an SU(5) group; the 15 observed left-handed quarks and leptons (excluding
left-handed antineutrinos) of each family are apportioned into $5$-dimensional
and $10^*$-dimensional representations of SU(5).  However, the pattern becomes
much simpler when SU(5) is included into the group SO(10) \cite{SO10,PS}.  The
$5$- and $10^*$- dimensional representations of SU(5) combine with an SU(5)
singlet, the right-handed neutrino, into a single $16^*$-dimensional spinor
representation of SO(10), a group of rank 5 whose representation members may be
identified by their coordinates in a 5-dimensional vector space.  The spinor
consists of next-to-nearest neighbors on the vertices of a 5-dimensional
hypercube in this space.  Its members may be identified by vectors of the form
\beq
\left( \pm \frac{1}{2}, \pm \frac{1}{2}, \pm \frac{1}{2}, \pm \frac{1}{2},
 \pm \frac{1}{2} \right)~,
\eeq
with an odd number of $+$ signs for $16^*$ and an even number for its conjugate
$16$ representation \cite{Slansky}. Other representations of SO(10) have simple
depictions in this language:  For example, members of the vector 10-plet of
SO(10) are denoted by
\beq
( \pm 1, 0, 0, 0, 0) + {\rm permutations}~.
\eeq
The group SO(10) has rank 5, so there are five mutually commuting observables
which may be represented in it.  As the color SU(3) subgroup of SO(10) has rank
two, one may take color isospin $I_{3C}$ and hypercharge $Y_C$ as two of the
observables.  For SU(2)$_L$ one takes its third component $I_{3L}$, while weak
hypercharge will be denoted by $Y_W$.  The electromagnetic charge is
$Q = I_{3L} + Y_W/2$.  A fifth
observable $Q_\chi$, lying in SO(10) but outside SU(5), will be defined shortly.

One may now measure the value of any observable for an SO(10) representation
member by taking its projection along a specific five-dimensional vector, e.g.:
$$
V(I_{3C}) = \left( +\frac{1}{2},-\frac{1}{2}, 0, 0, 0 \right)~;~~
V(Y_C)    = \left( +\frac{1}{3}, \frac{1}{3}, -\frac{2}{3}, 0, 0 \right)~,
$$
$$
V(I_{3L}) = \left( 0,0,0,+\frac{1}{2},-\frac{1}{2} \right)~;~~
V(Y_W)    = \left( -\frac{2}{3}, -\frac{2}{3}, -\frac{2}{3},1,1\right)~;~~
V(Q)      = \left( -\frac{1}{3}, -\frac{1}{3}, -\frac{1}{3},1,0 \right)~.
$$
An additional observable denoted can be represented by $V(Q_\chi) = (1,1,1,1,1)
/\sqrt{10}$, where we have chosen to normalize $Q_\chi$ in the same way as
$I_{3C}$ or $I_{3L}$.  In the 16-plet of SO(10), $5$-plet SU(5) members have
$Q_\chi = -3/\sqrt{40}$, $10^*$-plet members have $Q_\chi = 1/\sqrt{40}$, and
the SU(5) singlet has $Q_\chi = 5/\sqrt{40}$.

The specific members of the left-handed family (\ref{eqn:fam}) may be
denoted by the following spinors.  A subscript $1,2,3$ denotes the color
label; we display only one color of each quark.  We shall adopt the shorthand
$\pm$ for coordinates $\pm1/2$ \cite{WZ}.  We shall also put a vertical bar
between the first three indices, denoting color SU(3), and the last two,
denoting weak SU(2).  We then have
\beq
u_{L1} = (+--|+-)~;~d_{L1}=(+--|-+)~;~\nu_L = (+++|+-)~;~e^-_L = (+++|-+)~.
\eeq
Each of these is a weak doublet with $I_{3L} = \pm 1/2$, as the last two
indices are unequal.  The corresponding antiparticles, obtained by reversing
the signs of the first four indices, are
\beq
\bar u_{L1} = (-++|--)~;~\bar d_{L1}=(-++|++)~;~\bar \nu_L = (---|--)~;
~e^+_L = (---|++)~.
\eeq
The $\bar \nu_L$ has no charges within the Standard SU(3)$_C\otimes$SU(2)$_L
\otimes$U(1)$_{Y_W}$ Model; it is {\it sterile}.

\begin{figure}[h]
\begin{center}
\includegraphics[height=3in]{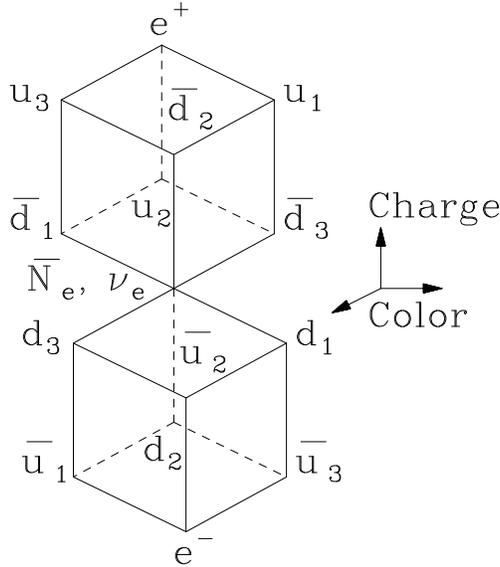}
\end{center}
\caption{Projection of SO(10) 16-plet describing a quark-lepton family into
the space of color (horizontal plane) $\otimes$ electric charge (vertical
axis)
\label{fig:cubes}}
\end{figure}

An interesting three-dimensional projection of the five-dimensional space may
be obtained by defining the horizontal plane to be the two-dimensional vector
space describing color SU(3) and the vertical axis to be electric charge.  The
members of an SO(10) 16-plet may then be represented as two cubes stacked
corner-to-corner, as shown in Fig.\ \ref{fig:cubes}.

So far we have not discussed the right-handed states.  These, it turns out, are
related to the corresponding left-handed states by a simple reversal of the
fifth index.  Thus, for right-handed particles we have
\beq
u_{R1} = (+--|++)~;~d_{R1}=(+--|--)~;~\nu_R = (+++|++)~;~e^-_R = (+++|--)~,
\eeq
while for right-handed antiparticles we have
\beq
\bar u_{R1} = (-++|-+)~;~\bar d_{R1}=(-++|+-)~;~ \bar \nu_R = (---|-+)~;
~e^+_R = (---|+-)~.
\eeq
It is now the right-handed {\it particles} which are electroweak singlets,
while the right-handed {\it antiparticles} are electroweak doublets.  In
particular, the right-handed neutrino $\nu_R$ is sterile with respect to
Standard Model charges.

All this is familiar to practitioners of grand unified theories.  Indeed,
the unbroken SO(10) symmetry is left-right symmetric \cite{PS,LRS}; it is
a non-zero expectation value of the charge $Q_\chi$ which destroys this
symmetry.  This could arise at any mass scale from a Higgs mechanism.  If
the scale is several TeV or less, one might be able to observe the
corresponding neutral gauge boson (a ``$Z_\chi$'' \cite{LRR,PL}) at the CERN
Large Hadron Collider (LHC).  A very large mass scale, however, could
be associated with a large Majorana mass of right-handed neutrinos.

One can also envision the breaking of SO(10) as proceeding first
through its subgroup SO(6) $\otimes$ SO(4) (easily illustrated on the fingers
of two hands).  The SO(6) is isomorphic to an SU(4) group which may be
thought of an extended color, regarding leptons as the fourth color \cite{PS}.
Its subgroup containing color is SU(3)$_C \otimes$ U(1)$_{B-L}$, where $B$ and
$L$ are baryon and lepton number.  The SO(4) is isomorphic to SU(2)$_L \otimes$
SU(2)$_R$.  The subsequent breaking of SU(2)$_R$ would be responsible for
parity-noninvariance of the electroweak theory.
A handy expression for electric charge, instead of the uninspiring
relation involving weak hypercharge, is $Q = I_{3L} + I_{3R} + (B-L)/2$.
The vectors projecting out $I_{3R}$ and $B-L$ are
\beq
V(I_{3R}) = \left( 0,0,0,+\frac{1}{2},\frac{1}{2} \right)~;~~
V(B-L) = \left( -\frac{2}{3}, -\frac{2}{3}, -\frac{2}{3},0,0 \right)~.
\eeq

What is puzzling about the parity operation is that, although it is a
transformation of the Poincar\'e group, it corresponds to a simple operation in
the SO(10) five-dimensional vector space.  Because it is a discrete
transformation, it evades the Coleman-Mandula theorem \cite{Coleman}
which forbids the combination of internal and Poincar\'e symmetries except as
a direct product.  Its violation is also deeply implicated in how the Standard
Model arises from some higher symmetry.  In the next section we argue that
such a symmetry is likely to exist on the basis of our very incomplete
knowledge about the nature of matter in the Universe.

\section{Expanded symmetries}

Ordinary matter makes up a small fraction of the known energy density of the
Universe; dark matter comprises about five times as much \cite{Komatsu:2008hk}.
We have little clue as to its nature.

Imagine a TeV-scale effective symmetry SU(3) $\otimes$ SU(2) $\otimes$ U(1)
$\otimes$ G, where the beyond-Standard-Model (BSM) group G could be any
number of extensions currently on the market.  One can classify the new
types of matter very generally as shown in Table \ref{tab:types}
\cite{Rosner:2005ec}:

\begin{table}[h]
\caption{Possible types of matter classified according to SM and BSM (G)
transformation.
\label{tab:types}}
\begin{center}
\begin{tabular}{c c c c} \hline \hline
Type of matter & Std.\ Model &    G    & Example(s) \\ \hline
Ordinary       & Non-singlet & Singlet & Quarks, leptons \\
Mixed          & Non-singlet & Non-singlet & Superpartners \\
Shadow         & Singlet     & Non-singlet & $E_8'$ of E$_8 \otimes$
E$_8'$ \\ \hline \hline
\end{tabular}
\end{center}
\end{table}

Grand unified theories well beyond SO(10) have been proposed.  The $E_8'$ of
E$_8 \otimes$ E$_8'$ in the heterotic string \cite{het} could play a role of
``shadow matter'' which communicates only weakly with our world.  The
fifth coordinate in the SO(10) description, whose reversal we have shown
induces parity reflections, could play a wider role in an extended vector
space of more than five dimensions.

The spinors of SO(2N) groups may be represented as alternate vertices of
hypercubes in $N$ dimensions.  These spinors and their conjugates each
have $2^{N-1}$ members.  If $N > 5$, one can ask what fraction of those
have the form $(+++++|a_{N-5} \ldots a_N)$ or $(-----|a_{N-5} \ldots a_N)$,
where the first five indices refer to the SO(10) subgroup of SO(2N),
and hence would be ``sterile'' under charges of the Standard Model.  The
answer is the same as for $N=5$, namely 1/16.  One is seeking, rather, a
scheme where {\it most} of the matter is sterile under Standard Model
charges.  The existence of a large amount of dark matter in our Universe
could be a key to guessing the structure of a large Grand Unified Group, and
perhaps incidentally helping to solve the mystery of parity violation.

\section{Conclusion} The advent of the CERN Large Hadron
Collider will offer one possible window into extended grand unified
theories, through the discovery of new forms of matter or new gauge bosons.
One of the simplest such examples would be the gauge boson $Z_\chi$ coupled
to the charge $Q_\chi$ mentioned above \cite{LRR,PL}.
It may turn out in retrospect that the role of parity and its violation in
our current understanding of unified theories was just a foretaste of a much
richer structure.

\section*{Acknowledgments}
I thank Z. Silagadze for alerting me to several interesting interesting papers
on mirror matter \cite{ZS}.
This work was supported in part by the United States Department
of Energy through Grant No.\ DE FG02 90ER40560.

\end{document}